\documentclass[a4paper,12pt]{article}
\usepackage{mathptmx}
\usepackage{epsfig}
\usepackage{float}
\usepackage{amssymb}
\usepackage{latexsym}
\usepackage{authblk}

\usepackage{amsfonts}
\newcommand{\beq}{\begin{equation}}
\newcommand{\eeq}{\end{equation}}
\newcommand{\beqa}{\begin{eqnarray}}
\newcommand{\eeqa}{\end{eqnarray}}

\usepackage{natbib}
\usepackage{graphicx}
\usepackage{multirow}
\begin{document}



\title{First analysis of GLE 72 event on 10 September 2017: Spectral and anisotropy characteristics}
\author[1]{A.L. Mishev}
\author[1,2]{I.G. Usoskin}
\author[3]{O. Raukunen}
\author[3]{M. Paassilta}
\author[3]{E. Valtonen}
\author[2]{L.G. Kocharov}
\author[3]{R. Vainio}
\affil[1]{Space Climate Research Unit, University of Oulu, Finland.}
\affil[2]{Sodankyl\"a Geophysical Observatory, University of Oulu, Finland.}
\affil[3]{Department of Physics and Astronomy, University of Turku, Finland.}
\maketitle
\begin{abstract}

Using neutron monitor and space-borne data we performed an analysis of the second ground level enhancement of solar cycle 24, namely the event of 10 September 2017 (GLE 72) and derive the spectral and angular characteristics of GLE particles. We employ new neutron monitor yield function and a recently proposed model based on optimization procedure. The method consists of simulation of particle propagation in a model magnetosphere in order to derive the cut-off rigidity and neutron monitor asymptotic directions. Subsequently the rigidity spectrum and anisotropy of GLE particles are obtained in their dynamical evolution during the event on the basis of inverse problem solution. The derived angular distribution and spectra are briefly discussed. 
\end{abstract}

 \normalsize

\label{cor}{\small For contact: alexander.mishev@oulu.fi}


\section{Introduction}

A detailed study of solar energetic particle (SEP) events provides important basis to understand their acceleration and propagation in the interplanetary space \citep{Deb88, Lockwood1990, Kallenrode1992, Reames99, Drake2009, Tylka2009, Li12, Vainio2013, Gopalswamy2014, Kocharov2017, Vainio2017}. Energetic and sporadic solar flares and coronal mass ejections (CMEs) can produce solar energetic particles \citep[e.g.][and references therein]{Reames99, Cliver04, Aschwanden12, Reames201353, Desai2016}. Normally the maximum energy of SEPs is several MeV/nucleon, rarely exceeding 100 MeV/nucleon. However, in some cases the SEP energy reaches several GeV/nucleon. While lower energy SEPs are absorbed in the atmosphere, those with energy above 300-400 MeV/nucleon generate an atmospheric shower, namely a complicated nuclear-electromagnetic-muon cascade consisting of large amount of secondary particles, which can reach the ground and eventually registered by ground based detectors, e.g. neutron monitors (NMs). The probability of occurrence of a high-energy SEP event is larger during the maximum and declining phase of the solar activity cycle \citep[\textit{e.g.}][]{Shea90}. This particular class of events is known as ground level enhancements (GLEs).

Such events can be conveniently studied using the worldwide NM network, a multi-instrumental equipment for continuous monitoring of the intensity of cosmic ray (CR) particles and registration of GLEs \citep{Sim53, Hat71, Bieber95, Stoker2000, Mav11, Gopalswamy201223, Moraal201285, Papaioannou2014423, Gopalswamy2014}. The distribution of NMs at different geographic regions allows one to obtain an exhaustive  record of cosmic rays in space, because their intensity is not uniform in the vicinity of Earth \citep{Bieber95, Mav11}. This fact plays an important role for GLE analysis, since an essential anisotropic part, normally during the event's onset is observed \citep{Vas06, But09, Mishev16SF}. The full list of NMs used for the analysis of GLE 72 on 10 September 2017, with the corresponding abbreviations, cut-off rigidities and altitudes is given in Table 1.

\begin{table*}[htp]
\caption{Neutron monitors with corresponding cut-off rigidities, geographic coordinates  and altitudes above the sea level used for the analysis of GLE 72.}             
\label{table:1}      
\centering   
\tiny    
\begin{tabular}{c c c c c }     
\hline\hline       

Station & latitude [deg] & Longitude [deg] &  $P_{c}$ [GV] & Altitude [m] \\ 
\hline                    
   Alma Aty (AATY) & 43.25  & 76.92  & 6.67 & 3340 \\  
   Apatity (APTY)& 67.55  & 33.33  & 0.48 &  177  \\
   Athens (ATHN) & 37.98  & 23.78  & 8.42 &  260  \\
   Baksan (BKSN) & 43.28  & 42.69  & 5.6  &  1700  \\ 
   Dome C (DOMC) & -75.06 & 123.20 & 0.1 & 3233   \\  
   Dourbes (DRBS) & 50.1  & 4.6    & 3.34 & 225    \\   
   Fort Smith (FSMT)  & 60.02 & 248.07  & 0.25 &  0  \\
   Inuvik (INVK)     & 68.35  & 226.28  & 0.16  &  21 \\
   Irkutsk (IRKT)  & 52.58 & 104.02 & 3.23 &  435       \\
   Jang Bogo(JNBG) & -74.37  & 164.13  & 0.1 & 29        \\ 
   Jungfraujoch (JUNG) & 46.55 & 7.98 & 4.46 & 3476      \\  
   Kerguelen (KERG) & -49.35 & 70.25 & 1.01 & 33          \\
   Lomnicky \v{S}tit (LMKS) & 49.2 & 20.22 & 3.72 & 2634    \\ 
   Magadan (MGDN) & 60.12 & 151.02 & 1.84 & 220            \\
   Mawson (MWSN) & -67.6 & 62.88 & 0.22 & 0               \\
   Mexico city (MXCO) & 19.33 & 260.8 & 7.59  & 2274              \\
   Moscow (MOSC) & 55.47 & 37.32 & 2.13 & 200              \\
   Nain (NAIN) & 56.55 & 298.32 & 0.28 & 0                      \\
   Newark (NWRK)& 39.70 & 284.30 & 1.97 & 50            \\
   Oulu (OULU) & 65.05 & 25.47 & 0.69 & 15                       \\
   Peawanuck (PWNK) & 54.98 & 274.56 & 0.16 & 52     \\
   Potchefstroom (PTFM) & -26.7 & 27.09 & 6.98 & 1351    \\
   Rome   (ROME)   & 41.9 & 12.52 & 6.11 & 60    \\
   South Pole (SOPO) & -90.00 & 0.0 & 0.01 & 2820             \\
   Terre Adelie (TERA) & -66.67 & 140.02 & 0 & 45            \\
   Thule (THUL) & 76.60 & 291.2 & 0.1 & 260                      \\
   Tixie Bay (TXBY) & 71.60 & 128.90 & 0.53 & 0                \\

\hline                  
\end{tabular}
\end{table*}

\begin{figure}
   \centering
   \includegraphics[width=0.5\textwidth]{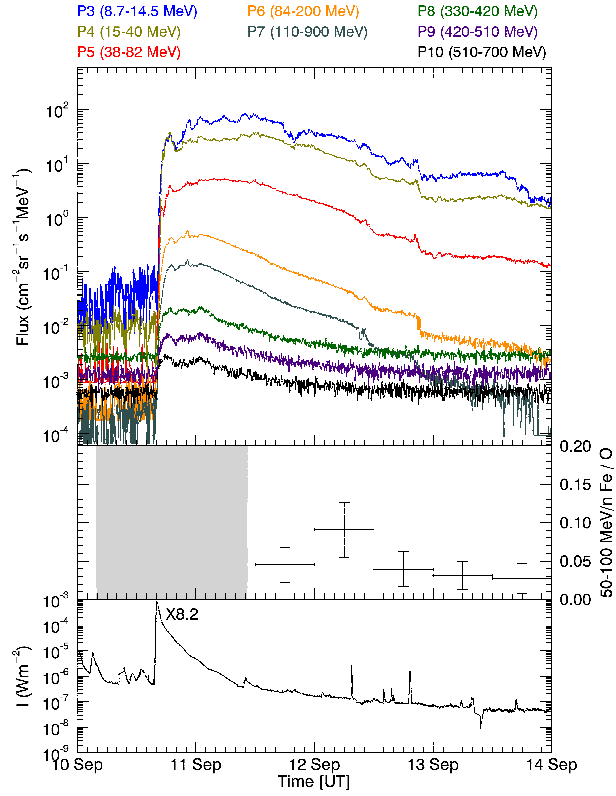}
      \caption{The 2017-09-10 SEP event and the related gradual soft X-ray flare as observed by GOES and SOHO / ERNE. The top panel shows the proton fluxes measured by GOES, the middle panel shows the Fe/O ratio at 50--100 MeV/n measured by SOHO / ERNE and the bottom panel shows the soft X-ray intensity measured by GOES. Note that SOHO / ERNE had a data gap in the beginning of the SEP event, indicated by the gray area in the respective panel.
              }
         \label{fig:SEPs}
   \end{figure}

The first ten days of September 2017 were characterized by intense solar activity. During this period several X-class flares and CMEs were produced. The GLE 72 on 10 September 2017 event was related to a X8.2 solar flare, a climax on a series of flares from active region 12673. It peaked at 16:06 UT, leading to a gradual solar energetic particle event measured by spacecraft up to proton energies exceeding 700 MeV/n (Fig.~\ref{fig:SEPs}), and to a very fast CME erupting over the west limb (Fig.~\ref{fig:CME}). The CME was first observed in SOHO / LASCO / C2 field of view at 16:00:07 UT. The initial speed of the CME was very high, 3620 km s$^{-1}$, as measured from the leading edge of the structure at position angle 270$^\circ$, i.e., to the west. The CME drove a shock front through the corona, with the flanks being traced in the atmospheric imaging assembly (AIA) of the Solar Dynamics Observatory (SDO) spacecraft image and the nose visible as a fainter structure in front of the brightest part of the CME. 

\begin{figure}
   \centering
   \includegraphics[width=0.8\textwidth]{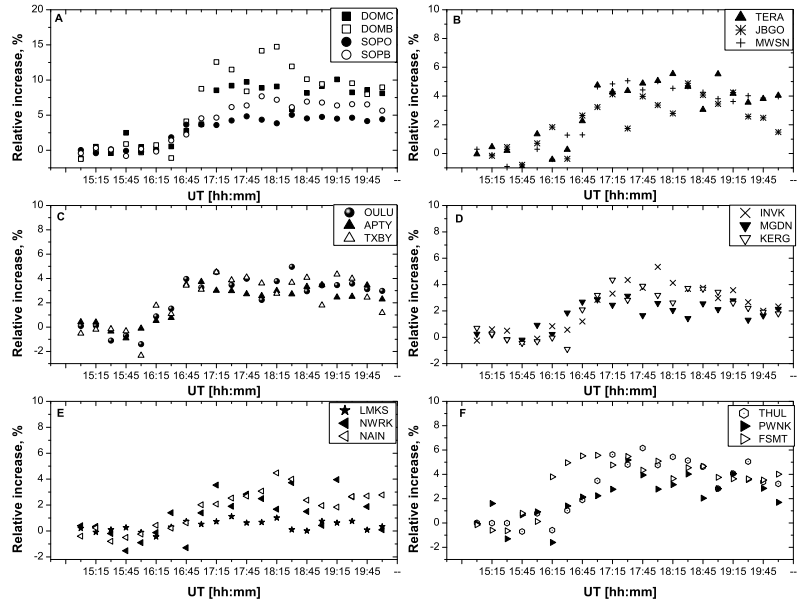}
      \caption{Count rate variation of NMs with statistically significant increase during GLE 72 on 10 September 2017. 
              }
         \label{Fig1}
   \end{figure}

The GLE onset was observed by several NM stations at about 16:15 UT (e.g. FSMT and  INVK with statistically significant increase) and the corresponding alert signal was revealed after 17:00 \citep{Souvatzoglou2014633}. However, a statistically significant and high enough signal, which allows one to derive the spectral and angular characteristics of SEPs with sufficient precision, was observed at 16:30 UT (see Sections 2 and 3). The strongest NM increases were observed at the DOMC/DOMB ($\approx$ 10--15 $\%$, Fig.~\ref{Fig1}a), SOPO/SOPB  ($\approx$ 5--8 $\%$, Fig.~\ref{Fig1}a) and FSMT ($\approx$ 6 $\%$, Fig.~\ref{Fig1}f) compared to the pre-increase levels, see details in Fig.~\ref{Fig1}. The count rate increase at FSMT was steeper compared to other stations, which recorded gradual increases. Herein DOMB, accordingly SOPB correspond to the lead free NMs at Dome C and South Pole stations, respectively. The event was characterized by a typical gradual increase, relatively hard rigidity spectrum and strong to moderate anisotropy during the event onset, which rapidly decreases resulting in a nearly isotropic flux. Similarly to some other events, it occurred during the recovery phase of a Forbush decrease, which is explicitly considered in our analysis. The background due to GCR is averaged over two hours before the event's onset, therefore we accounted for the corresponding variations of each NM count rates for exact computation of the background. Herein, we perform  an analysis of this event using 5-min integrated NM data retrieved from neutron monitor data base NMDB  (\texttt{http://www.nmdb.eu/}) \citep [\textit{e.g}.][]{Mav11}, available also at the international GLE database (\texttt{http://gle.oulu.fi/$\#$/}). 

\section{Modelling the neutron monitor response}
In this study we employ a method similar to  \citep{Shea82, Humble91, Cramp97, Bom06, Vas06, Vas08}. The detailed description of the method is given elsewhere \citep{Mishev14c, Mishev16SF, Mishev2017_icrc1, Mishev2017swsc}. The method implies modelling of NM response using an initial guess similarly to \citet{Mishev16SFE, Kocharov2017}, and/or to \citet{Cramp95} and optimization procedure over a selected space of unknowns describing SEP characteristics. Note, that for a good convergence of the optimization we need about $2(n-1)$ NM stations with non-null response, where $n$ is the number of unknowns  \citep[e.g.][]{Himmelblau72}. However, our method allows to derive a robust solution even in the case of relatively weak recorded NM increases on the basis of a specific numeric procedure, namely using a variable regularization  \citep[e.g.][]{Tikhonov1995, Mishev2017swsc}. 

Herein, we used a newly computed NM yield function, which is in a good agreement with experiments and recent modelling \citep{Mishev13b, Gil15, Mangeard201611}. We reduce model eventual uncertainties related to normalization of high-altitude NMs to sea level by employing yield function corresponding to the proper altitude of NMs above sea level \citep{Mishev2015_icrc1} and appropriate scaling of mini NM to a standard 6NM64  \citep{Caballero-Lopez20167461, Lara20161441}. 

In our model we can use as an approximation a modified power law or an exponential rigidity spectrum similarly to \citet{Cramp97, Vas08}.
The rigidity spectrum of SEPs described by a modified power is given by the expression:

\begin{equation}
    J_{||}(P)= J_{0}P^{-(\gamma+\delta\gamma(P-1))}
        \label{simp_eqn2}
   \end{equation}

\noindent where $J_{||}(P)$ denotes the flux of particles with rigidity $P$, which arrive from the Sun along the axis of symmetry identified by geographic latitude $\Psi$ and longitude $\Lambda$. The spectrum is described by $\gamma$ - power-law spectral exponent  and the rate of the spectrum steepening $\delta\gamma$. 

Accordingly, the exponential rigidity spectrum is given by:

\begin{equation}
    J_{||}(P)= J_{0}\exp(-P/P_{0}).
        \label{simp_eqn3}
   \end{equation}

\noindent where $P_{0}$ is a characteristic proton rigidity.

The pitch angle distribution in all cases is modelled as superposition of two Gaussians: 

\begin{equation}
        G(\alpha(P)) \sim \exp(-\alpha^{2}/\sigma_{1}^{2}) + B\cdot\exp(-(\alpha-\pi)^{2}/\sigma_{2}^{2}) 
        \label{simp_eqn4}
   \end{equation}

\noindent where $\alpha$ is the pitch angle, $\sigma_{1}$ and $\sigma_{2}$ quantify the width of the pitch angle distribution, B describes the amount of particle flux arriving from the anti-sun direction. This allows us to consider a  bidirectional particle flow.

The optimization is performed by minimizing the squared sum of difference between the modelled and measured NM responses employing the Levenberg-Marquardt method \citep{Lev44, Mar63} with variable regularization \citep{Tikhonov1995}, similarly to \citet{Mavrodiev2004359}. We assess the goodness of the fit on the basis of several criteria. The general criterion $\mathcal{D}$, i.e. the residual (Equation 4) is according to \citep[e.g.][]{Himmelblau72, Den83}.

\begin{equation}
\mathcal{D}=\frac{\sqrt{\sum_{i=1}^{m} \left[\left(\frac{\Delta N_{i}}{N_{i}}\right)_{mod.}-\left(\frac{\Delta N_{i}}{N_{i}}\right)_{meas.}\right]^{2}}}{\sum_{i=1}^{m} (\frac{\Delta N_{i}}{N_{i}})_{meas.}}
\label{simp_eqn6}
   \end{equation} 
   
According to our experience a good convergence of the optimization process and a robust solution are reached for $\mathcal{D}$ $\le$ 5 $\%$ similarly to \citet{Vas06, Mishev16SF}. Normally $\mathcal{D}$ is roughly 5 $\%$ for strong events and $\sim$ 10--15 $\%$ for weak events such as GLE 72 \citep[e.g.][]{Mishev2017_icrc1, Mishev2017swsc}. Therefore, we use additional criteria, namely the relative difference between the observed and calculated relative NM count rate increase should be about 10--20 $\%$ for each station and the residuals should have a nearly-symmetric distribution, viz. the number of NMs under and/or over estimating the count rate must be roughly equal \citep[e.g.][]{Himmelblau72}. 

The modelling of SEPs propagation in the geomagnetosphere used for computation of the cut-off rigidities and asymptotic directions of NMs \citep{Cook91} is performed with the MAGNETOCOSMICS code \citep{Desorgher05}, using International Geomagnetic Reference Field (IGRF) geomagnetic model (epoch 2015) as the internal field model \citep{Lan87} and the Tsyganenko 89 model as the external field \citep{Tsyganenko89}. This combination provides straightforward and precise modelling of SEPs propagation in the Earth's magnetosphere \citep{Kudela04, Kud08, Nevalainen13}.  

\section{Results of the analysis}
We studied different cases, assumed in our model, of spectral and PAD functional shapes, namely modified power law or exponential rigidity spectra of SEPs, single or double Gaussian PAD, which encompass all the possibilities in the model (Equations 2--4). An example of several computed asymptotic directions used for GLE 72 analysis is presented in Figure 3. Here we plot the asymptotic directions in the rigidity range  from 1 to 5 GV (for DOMC and SOPO respectively from  0.7 to 5 GV ) in order to show the range of maximal NM response, while in the analysis we used the range between the lower rigidity cut-off  of the station $P_{cut}$, i.e. the rigidity of the last allowed trajectory, below which all trajectories are forbidden, and the maximal assumed rigidity of SEPs being 20 GV. 

\begin{figure}[H]
   \centering
   \includegraphics[width=0.5\textwidth]{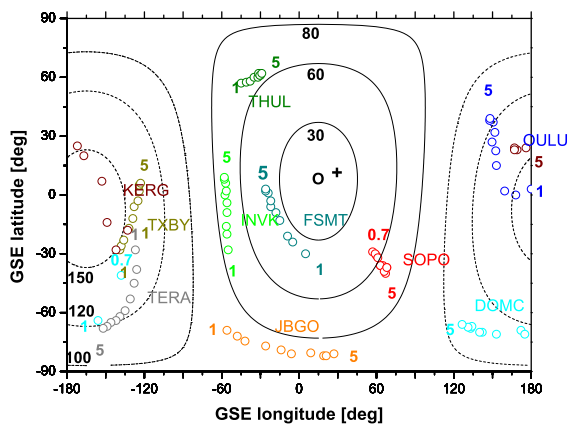}
      \caption{Asymptotic directions of several NM stations during the GLE 72 event on 10 September 2017 at 16:45 UT. The color lines and numbers indicate the NM stations and asymptotic directions (abbreviations given in Table 1). The small oval depicts the derived apparent source position, while the cross the IMF according to ACE satellite measurements. The lines of equal pitch angles relative to the derived anisotropy axis are plotted for 30$^{\circ}$, 60$^{\circ}$, 80$^{\circ}$ for sunward directions (solid lines), 100$^{\circ}$, 120$^{\circ}$  and 150$^{\circ}$, for anti-Sun direction (dashed lines). 
              }
         \label{Fig2}
   \end{figure}

The best fit is obtained assuming a modified power law rigidity spectra of SEPs and double Gaussian PAD (see below). For illustration we plot results with similar fit quality $\mathcal{D}$. The derived rigidity spectra of the high-energy SEPs during different stages of the event are presented in Figure 4, assuming a power law rigidity spectrum and a wide PAD fitted with a double Gaussian (Equations 1 and 3). 

\begin{figure}[H]
   \centering
   \includegraphics[width=0.8\textwidth]{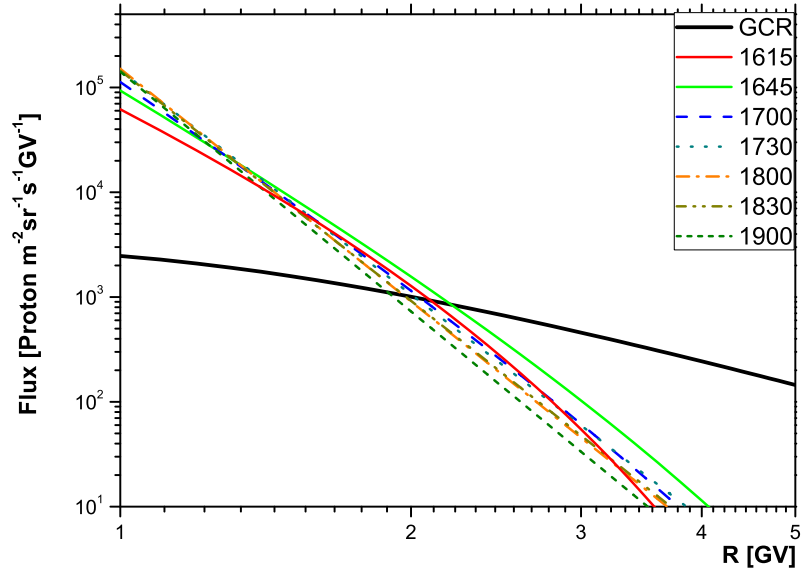}
      \caption{Derived SEP rigidity  spectra during GLE 72 on 10 September 2017. The black solid line denotes the GCR particle flux, which corresponds to the time period of the GLE 72 occurrence and includes protons and $\alpha$-particles, the latter representative also for heavy nuclei. Time (UT) corresponds to the start of the five minute interval over which the data are integrated. 
              }
         \label{Fig3}
   \end{figure}

The corresponding pitch angle distributions assuming a double Gaussian PAD are presented in Figure 5. 

\begin{figure}[H]
   \centering
   \includegraphics[width=0.8\textwidth]{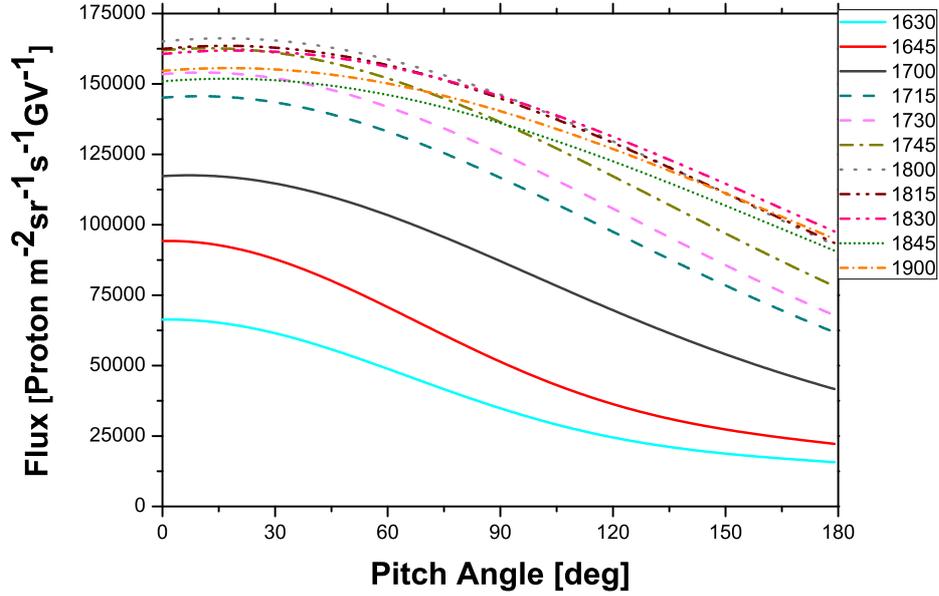}
      \caption{Derived PAD during GLE 72 on 10 September 2017.  Time (UT) corresponds to the start of the five minute interval over which the data are integrated. 
              }
         \label{Fig3}
   \end{figure}

An analysis was performed assuming a single Gaussian PAD. The derived rigidity spectra appear with similar slope. The corresponding particle flux can be adjusted on the basis of a normalization. This case results in greater residuals $\mathcal{D}$. Moreover, the two additional criteria for the fit goodness, namely a nearly-symmetric distribution of the residuals and relative difference  between the observed and calculated NM increases to be in the order of 10--20 $\%$ is not achieved, specifically during the initial phase of the event.
 
Figure 6 presents the derived SEP rigidity spectra during different stages of the event, assuming a single Gaussian PAD. 

\begin{figure}[H]
   \centering
   \includegraphics[width=0.8\textwidth]{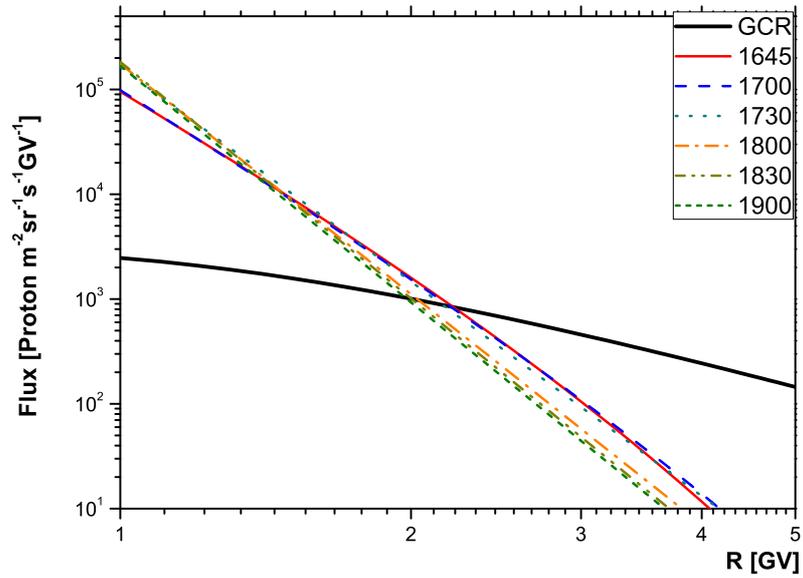}
      \caption{Derived SEP rigidity  spectra during GLE 72 on 10 September 2017. The black solid line denotes the GCR particle flux, which corresponds to the time period of the GLE 72 occurrence and includes protons and $\alpha$-particles, the latter representative also for heavy nuclei. Time (UT) corresponds to the start of the five minute interval over which the data are integrated. 
              }
         \label{Fig3}
   \end{figure}

The corresponding PADs assuming a single Gaussian are presented in Figure 7. 

\begin{figure}[H]
   \centering
   \includegraphics[width=0.8\textwidth]{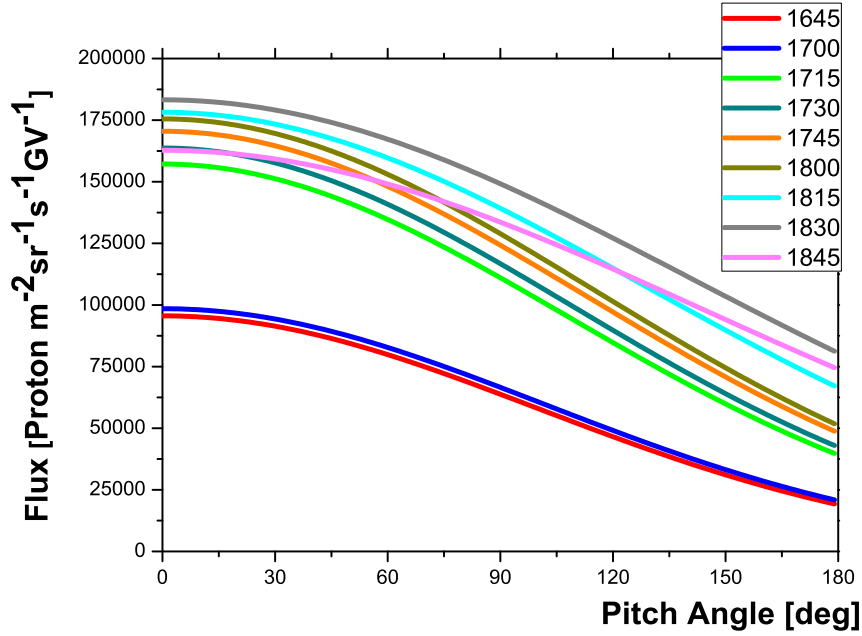}
      \caption{Derived PAD during GLE 72 on 10 September 2017.  Time (UT) corresponds to the start of the five minute interval over which the data are integrated. 
              }
         \label{Fig3}
   \end{figure}

Finally, we tried  to fit the global NM network response assuming an exponential rigidity spectrum of SEPs (Equation 2). In this case, the residual $\mathcal{D}$ is considerably greater compared to previous cases, namely $\mathcal{D}$ is  $\approx$ 40--50 $\%$. 

The accuracy of the modelling is shown by a comparison between the modelled and observed responses for several NMs during the GLE 72 on 10 September 2017 (Figure 8). Note, that the quality of the modelling is similar for the other NM stations. The comparison is performed in the case of double Gaussian PAD, while in the case of simple Gaussian the difference between modelled and experimental NM responses is considerably greater.

\begin{figure}[H]
   \centering
   \includegraphics[width=\textwidth]{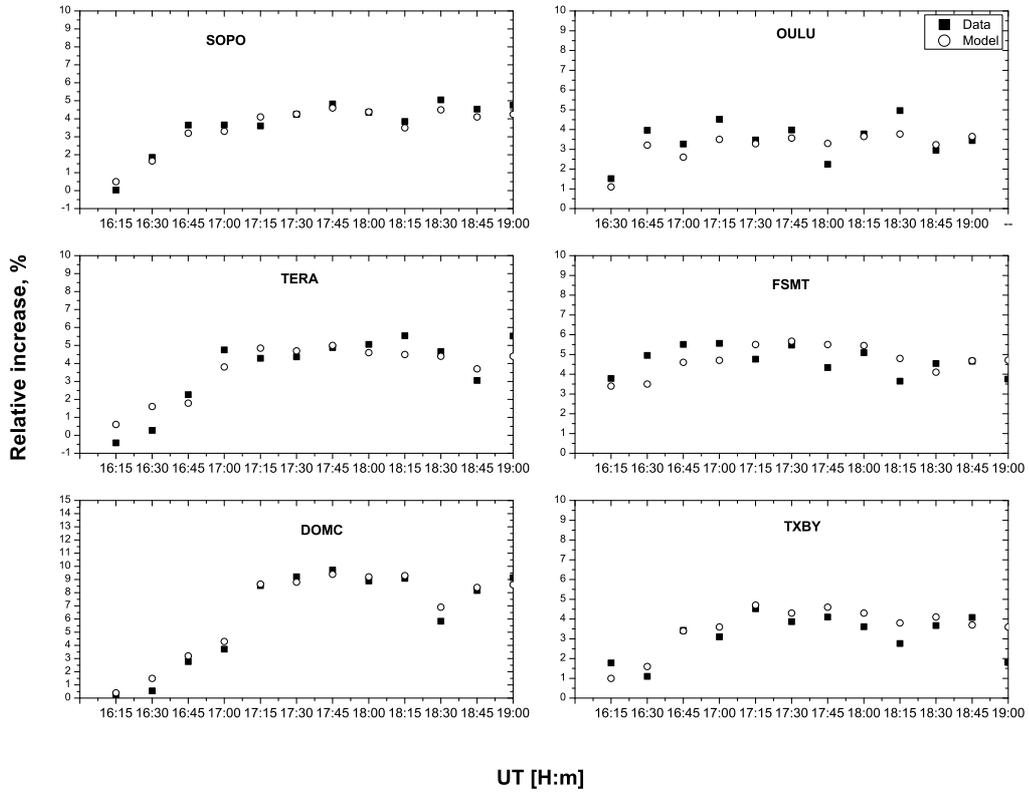}
      \caption{Modelled and observed responses of several NM stations during the GLE 72 on 10 September 2017. The accuracy of the fit for other stations is of the same order.
              }
         \label{Fig7}
   \end{figure}

The particle fluence (energy, time and angle integrated particle flux) of the GLE 72 is presented in Figure 9 for the early and late phase of the event respectively. Accordingly the normalized to one hour  SEP fluence is shown in Figure 10. As expected during the early phase of the event, the fluence is dominated by the high-energy part of SEPs.  Accordingly during the late phase of the event the SEP flux increased at low energies.  

According to our analysis the related SEPs had the Fe/O abundance ratio below 0.1 at 50-100 MeV/n (Fig.3), which is typical for gradual SEP events \citep[e.g.][and references therein]{Desai2008, Desai2016}. 

\begin{figure}[H]
   \centering
   \includegraphics[width=\textwidth]{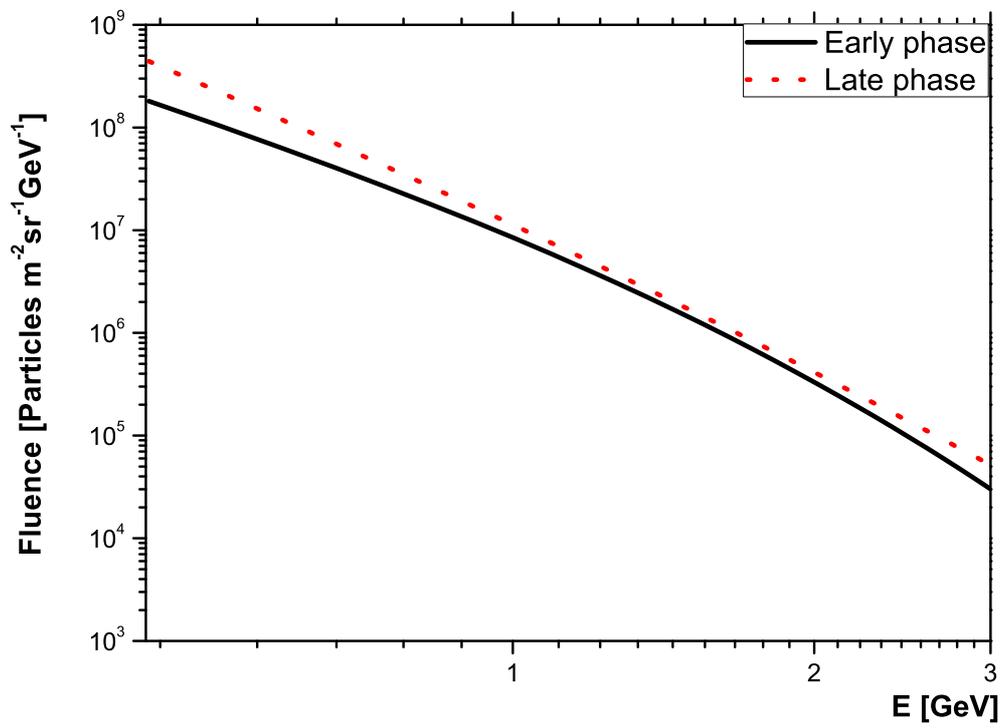}
      \caption{Fluence of high-energy SEP during GLE 72. The SEPs fluence for the early phase is angle integrated from 16:15 to 17:15 UT, while the fluence for the late phase is angle integrated from 17:15 to 19:15 UT.
              }
         \label{Fig8}
   \end{figure}

\begin{figure}[H]
   \centering
   \includegraphics[width=\textwidth]{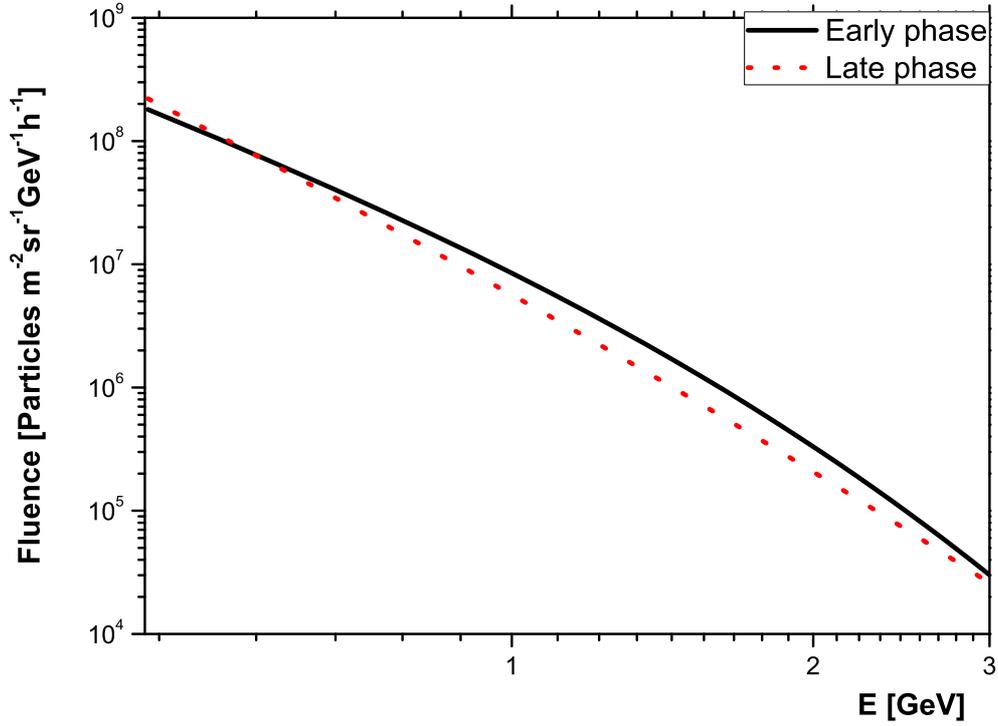}
      \caption{Normalized to 1h fluence of high-energy SEP during GLE 72 for the early and late phase of event respectively.
              }
         \label{Fig9}
   \end{figure}

\section{Discussion}
Here we derived the spectral and anisotropy characteristics of SEPs during the weak GLE 72 event on 10 September 2017 using NM data. According to our analysis, the apparent source position was close to the direction of the IMF lines (Figure 3), the latter being estimated from the ACE satellite measurements explicitly considering the time shift of the field direction at the nose of the Earth's bow shock similarly to \citet{Mishev2017swsc}. According to our estimations the uncertainty of the derived apparent source position is about 10--15 degrees. This implies that particles were propagating from the Sun close to the nominal Parker spiral.

The best fit of the modelled global NM responses was achieved assuming a modified power law rigidity spectrum (Eq.1) and a wide PAD fitted with double Gaussian (Eq.3). The SEP spectrum was moderately hard during the event onset and constantly softened throughout the event. A marginal hardening was observed at about 18:15 UT. A steepening of the spectrum  with rigidity, was observed during the initial phase of the event, which vanished lately. Hence, after 17:15 UT a pure power law rigidity spectrum was derived. An important anisotropy during the event onset was observed, since there were statistically significant responses at FSMT and INVK NM stations, but no or marginal at other stations. These NM stations responded to SEP fluxes with narrow pitch angles as their asymptotic cones were close to the direction of the IMF lines. The angular distribution of the SEPs during the event broadened out throughout the event. The timing of the event (first increase at 16:15 with a clear shock formed already at 16:00 in the corona) is consistent with a hypothesis of particle acceleration at a coronal shock driven by the CME. In addition, the ratio Fe/O at 50--100 MeV/n is low, which is also  consistent with gradual event \citep{Desai2008, Desai2016}. However, a more detailed and deep analysis is necessary using all the available data in order to derive relevant information about the SEP acceleration \citep[e.g.][]{Kocharov2017}.

This derived PAD could be a result of a focused transport of SEPs with an enhanced isotropization 
or alternatively we can speculate that may be due to a non-standard mode of the particle propagation caused by an interplanetary magnetic field structure associated with previous CME \citep{Ruffolo06}. A detailed analysis and modelling similarly to \citet{Kocharov2017} is planned for forthcoming work. 

\section{Conclusions}
Here, we have employed an improved method, compared to \citep{Mishev14c} of an analysis of data from the global neutron monitor network, namely the response of each NM is computed using a yield function corresponding to the exact NM altitude a.s.l. and self-consistent and robust optimization similarly to \citet{Mishev2017swsc}, applied for the GLE 72 on 10 September 2017. The data set included records from  27 NMs distributed over the globe, which encompass a wide range of the particle arrival directions and rigidities. 

The method consists of consecutive computation of asymptotic directions and cut-off rigidity of the NM stations; modelling of the NM responses and solution of inverse problem. The modelling of the NMs responses is performed applying Tsyganenko 1989 and IGRF magnetospheric field models and using new NM yield function. Herein, the method employs a modified power law or exponential rigidity spectrum of SEP and superposition of two Gaussians for PAD. 

We have studied several possible cases of spectral and angular distribution of SEPs, namely exponential or power law rigidity spectrum and single or double Gaussian PAD. Hence, we derived the spectral and angular characteristics of GLE particles. The time evolution of the spectral and angular characteristics is derived in the course of the event (see the supplementary materials, namely animations, which demonstrate the rigidity spectra and PAD evolution throughout the event). The 10 September 2017 event has revealed  a wide PAD, best fitted with a double Gaussian, except the event onset phase, when a narrow angular distribution of SEPs is derived. The PAD parameters depict one maximum of SEP flux at/or near zero pitch angle. This is qualitatively consistent with hypothesis of focused transport \citep[e.g.][]{Agueda2012}. Fit with a single Gaussian PAD was excluded during the analysis, specifically during the initial phase of the event (16:30 UT), but the event onset with narrow PAD. The two possible fits are compared and their quality was briefly discussed.

The best fit of spectral characteristics of SEPs corresponds to a modified power law rigidity spectrum. The rigidity spectrum during the event onset is harder than that during the late phase of the event. From the derived spectra and PAD alone it is hardly possible to define an exact scenario of particle acceleration and transport, but the timing of the event as well as the derived ratio Fe/O at 50--100 MeV/n is qualitatively consistent with a shock acceleration. This study gives a basis for subsequent studies of SEP acceleration and interplanetary transport. 

\section*{Acknowledgements}
This work was supported by the Academy of Finland (project No. 272157, Center of Excellence ReSoLVE and project No. 267186). Operation of the DOMC/DOMB NM was possible due to support of the French-Italian Concordia Station (IPEV program n903 and PNRA Project LTCPAA PNRA14 00091), projects CRIPA and CRIPA-X No. 304435 and Finnish Antarctic Research Program (FINNARP). We acknowledge NMDB and all the colleagues and PIs from the neutron monitor stations, who kindly provided the data used in this analysis, namely: Alma Ata, Apatity, Athens, Baksan, Dome C, Dourbes, Forth Smith, Inuvik, Irkutsk, Jang Bogo, Jungfraujoch, Kerguelen, Lomnicky \v{S}tit,  Magadan, Mawson, Mexico city, Moscow, Nain, Newark, Oulu, Peawanuck, Potchefstroom, Rome, South Pole, Terre Adelie, Thule, Tixie.




\end{document}